\documentclass[a4paper]{article}
\usepackage{authblk} % For multiple authors with different affiliations

\usepackage[english]{babel}
\usepackage[utf8x]{inputenc}
\usepackage[T1]{fontenc}

\usepackage{algpseudocode}
\usepackage{amsmath}
\usepackage{multirow}
\usepackage{subcaption}
\usepackage{float}     
\usepackage[numbers]{natbib}
 \usepackage{graphicx}
\usepackage{amssymb}
\usepackage{tcolorbox}
\usepackage[a4paper,top=3cm,bottom=2cm,left=3cm,right=3cm,marginparwidth=1.75cm]{geometry}

\usepackage{amsmath}
\usepackage{verbatim}
\usepackage{graphicx}
\usepackage{diagbox}
\usepackage{enumitem}
\usepackage{wrapfig}
\usepackage{multirow}
\usepackage{indentfirst}
\usepackage{tabularx}
\usepackage{booktabs}
\usepackage{ulem}
\usepackage[colorlinks=true, allcolors=blue]{hyperref}
\newtcolorbox[number within=section]{promptbox}[2][]{%
  colback=blue!5,
  colframe=blue!40,
  coltitle=black,
  fonttitle=\bfseries,
  title=~\thetcbcounter #2,
  #1
}

\title{Overview of TREC 2025 Biomedical Generative Retrieval (BioGen) Track
}

\author[1]{Deepak Gupta}
\author[1]{Dina Demner-Fushman}
\author[2]{William Hersh}
\author[2]{Steven Bedrick}
\author[3]{Kirk Roberts}
\affil[1]{National Library of Medicine, NIH}
\affil[2]{Oregon Health \& Science University}
\affil[3]{UTHealth Houston}

\date{}
\begin{document}
\maketitle
\section{Overview} \label{sec:into}
Recent advances in large language models (LLMs) have made significant progress across multiple biomedical tasks, including biomedical question answering, lay-language summarization of the biomedical literature, and clinical note summarization. These models have demonstrated strong capabilities in processing and synthesizing complex biomedical information and in generating fluent, human-like responses. Despite these advancements, hallucinations or confabulations remain key challenges when using LLMs in biomedical and other high-stakes domains. 
Inaccuracies may be particularly harmful in high-risk situations, such as medical question answering \cite{zhao2024heterogeneous}, making clinical decisions, or appraising biomedical research. Studies on evaluation of the LLMs’ abilities to ground generated statements in verifiable sources have shown that models perform significantly worse on lay-user generated questions \cite{wu2024well}, and often fail to reference relevant sources\cite{basaragin-etal-2024-know}. This can be problematic when those seeking information seek evidence from studies to support LLM claims \cite{jamiaocae014}. Unsupported statements are a major barrier to using LLMs in any applications that may affect health. Methods for grounding generated statements in reliable sources, along with practical evaluation approaches, are needed to overcome this barrier.  To this end, the TREC 2024 BioGen track introduced reference attribution to mitigate the generation of false statements by LLMs answering biomedical questions. The TREC 2025 BioGen track\footnote{\url{trec-biogen.github.io/docs/}} continues the reference attribution task with an additional task: grounding the answer. This requires citing references to support the text of the sentences and the overall answer from the LLM output for each topic.

The track received a total of $87$ runs\footnote{The term \textit{runs} and \textit{submissions} are used interchangeably in this paper.} from $15$ different teams. For task A, participants proposed a variety of approaches to retrieve the supports/contradictions of the statements, including natural language inference (NLI) and prompting an LLM to identify supporting or contradictory references. For task B, most runs used a two-stage retrieval-augmented generation (RAG) approach to generate answers with references. In the first stage, a retriever was used to identify relevant literature, and in the second stage, LLMs were used to generate answers with appropriate references.

\section{Tasks} \label{sec:task}
\begin{enumerate}
\item \textbf{Task A (Grounding Answer)}: Given a biomedical question, a stable version of PubMed documents, and an answer sentence to the question, the task is to ground each sentence of the answer with appropriate PubMed documents by providing their PMIDs.
For each answer sentence, participants were also provided with slightly outdated supporting PMIDs.
The system-generated PMIDs should include additional relevant documents. Since identifying contradictory references is crucial in the biomedical domain, systems were expected to provide PMIDs that contradict each answer sentence.
This task serves as a foundational step, preparing participants to effectively tackle the more complex task of Reference Attribution (Task B). For each answer sentence, the  PMIDs returned by the participants should meet the following requirements:
\begin{enumerate}
    \item The supporting PMIDs should be provided in addition to the existing supporting PMIDs already provided with each answer sentence.
    \item There should be no more than three new PMIDs per answer sentence for both supporting and contradicting assertions. Contradictory assertions are more important; if the system identifies both supporting and contradictory PMIDs, present the contradictory ones first.
    \item The PMIDs must be selected only from the valid set of PubMed articles released with the dataset.
\end{enumerate}

    \item 
\textbf{Task B (Reference Attribution):} Given a biomedical topic (question) and a stable version of PubMed documents. The task is to generate answers that include LLM output and attributions (cited references from PubMed) for each answer sentence. The generated answer must meet the following requirements:
\begin{enumerate}
 \item  The total length of the generated answer should be within 250 words.
 \item Each sentence\footnote{assertions/statements were approximated as sentences.} must be supported by up to three attributions (cited references).
 \item The PMIDs must be selected only from the valid set of PubMed articles released with the dataset.
 \item Each document should be referenced in the answers as PMIDs provided in the list with each answer sentence. 
\end{enumerate}
The participants were allowed to submit up to ten (10) runs for each task.
\end{enumerate}

\begin{table}[]
\resizebox{\columnwidth}{!}{%
\begin{tabular}{l|l}
\hline
\textbf{Question} & \textit{How can I lower my cortisol level?
} \\ 
\textbf{Answer Sentence} & \begin{tabular}[c]{@{}l@{}}\textit{Stress management, such as mindfulness, meditation, and relaxation, or } \\  \textit{interacting with dogs for an hour, can lower cortisol levels.}\end{tabular} \\ 
\textbf{Supported Citations} &\textit{ 35989236,
                    37879237
} \\ \hline
 &  \\ \hline
\textbf{Question} & \textit{How can I lower my cortisol level?
} \\ 
\textbf{Answer Sentence} & \begin{tabular}[c]{@{}l@{}}\textit{melatonin  and dehydroepiandrosterone (DHEA)  might help as well.} \end{tabular}\\ 
\textbf{Supported Citations} &\textit{ 34342920,
                    34616464
} \\  \hline
 &  \\ \hline
\textbf{Question} & \textit{How does CBD effect liver enzymes?
} \\ 
\textbf{Answer Sentence} & \begin{tabular}[c]{@{}l@{}}\textit{Cannabidiol (CBD) is a non-psychoactive (not causing 'high') chemical found} \\ \textit{in Cannabis (marijuana).} \end{tabular}\\ 
\textbf{Supported Citations} &\textit{ 338397045
} 
 \\ \hline
\end{tabular}%
}
\caption{Example topics from the Task A of the BioGen track.}
\label{tab:example-topic-topic-a}
\end{table}

\begin{table}[]
\resizebox{\columnwidth}{!}{%
\begin{tabularx}{\columnwidth}{lX}
\toprule
\textbf{Topic} & \textit{UTI diagnostic tests} \\
\textbf{Question} & \textit{What tests might be done to verify that I have a UTI?} \\
\textbf{Narrative} & 
\textit{A patient with discomfort during urination and flank pain has been on treatment for a urinary tract infection for three days without improvement and is asking which tests can definitively confirm the infection.} \\
\midrule
\textbf{Topic} & \textit{Hospitalization for sepsis} \\
\textbf{Question} & 
\textit{If you have sepsis with no breathing or gastrointestinal symptoms but experience extreme thirst, does this require hospitalization?} \\
\textbf{Narrative} & 
\textit{The provider suspects sepsis and recommends hospitalization for confirmation and treatment, but the patient, facing urgent deadlines, questions whether admission is necessary given the absence of respiratory or gastrointestinal symptoms.} \\
\midrule
\textbf{Topic} & \textit{Malocclusion related to tonsils and adenoids} \\
\textbf{Question} & 
\textit{What are tonsils and adenoids, and how are they related to malocclusion?} \\
\textbf{Narrative} & 
\textit{Parents of an eight-year-old girl were informed that she has enlarged tonsils and adenoids and was diagnosed with Class II malocclusion, leading them to question whether the two conditions are related.} \\
\bottomrule
\end{tabularx}
}
\caption{Example topics from the Task B of the BioGen track.}
\label{tab:example-topic-topic-b}
\end{table}

\begin{figure}[!t]

\fbox{\begin{minipage}{\textwidth}
\noindent \textbf{Topic:} \textit{Malocclusion related to tonsils and adenoids} \\
\textbf{Question:} \textit{What are tonsils and adenoids, and how are they related to malocclusion?} \\
\noindent \textbf{Narrative}: \textit{The patient is interested in the link between iron and infection, the role iron plays in infection and the implications for the COVID-19 course.} \\
\noindent \textbf{Sample Answer [adapted to patient-level health literacy]}: \textit{The tonsils and adenoids are part of the lymphatic system, located in the throat} [10859465, 30969614, 37512798]. \textit{As part of the immune system, they help fight infection in the upper respiratory tract} [10859465, 30969614, 37512798]. \textit{In their role of immunological sentinels, the tonsils and adenoids are constantly exposed to infections, which may cause their enlargement and inflammatory conditions such as tonsillitis} [30969614, 37512798].  \textit{Enlarged tonsils and adenoids can lead to habitual mouth breathing or cause obstructive sleep apnea} [12474699, 8054305, 2591488]. \textit{Mouth breathing may result in altered development of the shape of the jaws and face} [12474699, 34734579]. \textit{Most studies show that children with enlarged tonsils and adenoids are more likely to have Class II or III malocclusion, depending on the location of the enlarged tonsils} [38145836, 21079167]. \textit{Another study concluded that the size of the tonsils is not a risk factor for malocclusion} [19282036].  \\

\end{minipage}}
\caption{Sample reference answer for one of the topics of Task B of the BioGen track}
 \label{fig:topic}
\end{figure}
\section{Topics}
\begin{enumerate}
    \item \textbf{Task A (Grounding Answer):} We construct\textcolor{blue}{ed} the topic (question, answer sentence, and corresponding supported citations) for Task A by leveraging the expert-curated answer and assessed supported citations from the BioGen 2024 task of Reference Attribution. In total, the topics include 40 questions, 194 answer sentences, and their corresponding supporting citations. Example topics are shown in Table \ref{tab:example-topic-topic-a}.
     \item \textbf{Task B (Reference Attribution):} We constructed $30$ BioGen 2025 topics that were developed using information requests submitted by self-identified non-clinicians to the National Library of Medicine. Example topics are shown in Table \ref{tab:example-topic-topic-b}.
\end{enumerate}

% \usepackage{graphicx}

\section{Data}
The BioGen task used the latest annual baseline snapshot of MEDLINE/PubMed, which goes approximately through the end of 2024. We provided a pre-processed set of $26,805,982$ PMIDs representing the abstracts in the 2024 snapshot. Participants were asked to cite and use the PMIDs when generating the answers in this collection. We also released\footnote{\url{http://bionlp.nlm.nih.gov/biogen-2025-document-collection.zip}} the indexed PubMed collection via PySerini\footnote{\url{https://github.com/castorini/pyserini}}. 
\begin{table}[h]
\centering
\resizebox{\columnwidth}{!}{%
\begin{tabular}{ll|l|lllll}
\hline
\multicolumn{2}{c|}{\textbf{Task A}} &  & \multicolumn{5}{c}{\textbf{Task B}}                                                                                       \\ 
\multicolumn{2}{c|}{\textbf{Submissions}} &
   &
  \multicolumn{2}{c}{\textbf{Early Submissions}} &
  \multicolumn{1}{l}{\textbf{}} &
  \multicolumn{2}{c}{\textbf{Final Submissions}} \\ \hline
\multicolumn{1}{l|}{\textbf{Team Name}} &
 \textbf{ \#Runs} &
   &
  \multicolumn{1}{l|}{\textbf{Team Name}} &
  \multicolumn{1}{l|}{\textbf{\#Runs}} &
  \multicolumn{1}{l|}{} &
  \multicolumn{1}{l|}{\textbf{Team Name}} &
  \textbf{\#Runs} \\ \hline
\multicolumn{1}{l|}{CLaC}       & 4  &  & \multicolumn{1}{l|}{hltcoe-rerank}   & \multicolumn{1}{l|}{10} & \multicolumn{1}{l|}{} & \multicolumn{1}{l|}{CLaC Lab} & 5 \\ 
\multicolumn{1}{l|}{dal}        & 2  &  & \multicolumn{1}{l|}{UAmsterdam}      & \multicolumn{1}{l|}{5}  & \multicolumn{1}{l|}{} & \multicolumn{1}{l|}{CLaC}     & 4 \\
\multicolumn{1}{l|}{GEHC-HTIC}  & 1  &  & \multicolumn{1}{l|}{GEHC}            & \multicolumn{1}{l|}{1}  & \multicolumn{1}{l|}{} & \multicolumn{1}{l|}{SIB}      & 4 \\ 
\multicolumn{1}{l|}{InfoLab}    & 5  &  & \multicolumn{1}{l|}{dal}             & \multicolumn{1}{l|}{8}  & \multicolumn{1}{l|}{} & \multicolumn{1}{l|}{InfoLab}  & 4 \\ 
\multicolumn{1}{l|}{polito}     & 1  &  & \multicolumn{1}{l|}{EvalHLTCOE}      & \multicolumn{1}{l|}{2}  & \multicolumn{1}{l|}{} & \multicolumn{1}{l|}{h2oloo}   & 4 \\ 
\multicolumn{1}{l|}{SIB}        & 7  &  & \multicolumn{1}{l|}{uniud}           & \multicolumn{1}{l|}{10} & \multicolumn{1}{l|}{} & \multicolumn{1}{l|}{GEHC}     & 1 \\
\multicolumn{1}{l|}{uniud}      & 4  &  & \multicolumn{1}{l|}{hltcoe-multiagt} & \multicolumn{1}{l|}{3}  & \multicolumn{1}{l|}{} & \multicolumn{1}{l|}{dal}         & 1  \\ 
\multicolumn{1}{l|}{}           &    &  & \multicolumn{1}{l|}{UDInfo}          & \multicolumn{1}{l|}{1}  & \multicolumn{1}{l|}{} & \multicolumn{1}{l|}{}         &   \\ \hline \hline
\multicolumn{1}{l|}{\textbf{7 Teams}} &
  \textbf{24 Submissions} &
  \textbf{} &
  \multicolumn{1}{l|}{\textbf{8 Teams}} &
  \multicolumn{1}{l|}{\textbf{40 Submissions}} &
  \multicolumn{1}{l|}{\textbf{}} &
  \multicolumn{1}{l|}{\textbf{7 Teams}} &
  \textbf{23 Submissions} \\ \hline
\end{tabular}%
}
\caption{Statistics of the team and their runs on each task of the BioGen 2025 track.}
\label{tab:team_participations}
\end{table}

% \usepackage{graphicx}
\begin{table}[]
\centering
\resizebox{0.8\columnwidth}{!}{%
\begin{tabular}{l|lccc}
\hline
Team Name       & \#Runs & Expert Evaluations & Complete & Incomplete \\ \hline \hline
hltcoe-rerank   & 10     & 6                           & 4        & 2          \\ 
UAmsterdam      & 5      & 5                           & 2        & 3          \\ 
GEHC            & 1      & 1                           & 1        & 0          \\
dal             & 8      & 6                           & 2        & 4          \\ 
EvalHLTCOE      & 2      & 2                           & 1        & 1          \\ 
uniud           & 10     & 5                           & 3        & 2          \\ 
hltcoe-multiagt & 3      & 3                           & 2        & 1          \\ 
UDInfo          & 1      & 1                           & 1        & 0          \\ 
Baseline        & 1      & 1                           & 1        & 0          \\  \hline \hline
Total           & 41     & 30                          & 17       & 13         \\ \hline 
\end{tabular}%
}
\caption{The statistics of the runs submitted in early submission, along with the number of runs chosen for expert evaluation. The \textbf{Complete} columns show, based on the selected expert evaluation, how many runs yielded the full assessment for each topic, the answer sentence, and the cited sources listed in those runs. }
\label{tab:expert-evaluations}
\end{table}

\section{Participating Teams and Submissions}
We used the NIST-provided Evalbase platform\footnote{\url{https://ir.nist.gov/evalbase}} to release the
datasets, registration, and submissions of the participating teams.
In total, $15$ teams participated in the BioGen track and submitted $87$ individual runs for the tasks. The team details, along with their submissions, are provided in Table \ref{tab:team_participations}.
For Task B, we encourage\textcolor{blue}{d} early submissions to obtain the expert evaluation of the submitted runs.
We received 40 submissions from 8 teams during the early submission period and 23 submissions from 7 teams during the final submission period.
Due to resource constraints, we select\textcolor{blue}{ed} 30 of the highest-priority runs (team-wise) for expert evaluation. The details are provided in Table \ref{tab:expert-evaluations}. 

\section{Baseline Approaches}

\begin{enumerate}
    \item \textbf{Task A (Grounding Answer):} We developed an NLI-based baseline system for Task~A. Toward this, for a given answer sentence, we first retrieved the relevant PubMed document using Pyserini from the indexed PubMed corpus. The top three-ranked documents are selected as supporting evidence. If any of these documents are already in the supported list, they are replaced by the next-highest-ranked documents until three unique supporting documents are obtained. To identify contradictory documents, each retrieved document is assessed using SciFive \cite{phan2021scifive}, a MedNLI-trained natural language inference model, which classifies document–answer sentence pairs as entailing, neutral, or contradictory.
    
    \item \textbf{Task B (Reference Attribution):} 
    We adopted an RAG-based approach, first fine-tuning a LLaMA2-7B \cite{touvron2023llama} model to generate answers with citations. We created the dataset for inference tuning from the PLABA \cite{attal2023dataset} collection, which includes questions and relevant documents, and CHQ-Summ \cite{yadav2022chq}, which contains consumer health questions. For CHQ-Summ, we first retrieve the most relevant documents to the question using Pyserini from the indexed PubMed corpus. Thereafter, we re-ranked the documents using GraphMonoT5 \cite{gupta2024empowering} re-ranker and selected the top 10 documents. Given the retrieved document, we use the bottom-line extractor \cite{demner2007answering} to extract 3 sentences as an extractive summary. The question, along with a summary of the top 10 documents, is passed to the GPT-3.5 Turbo model to generate the answer and cite the statements using the prompt \cite{gao2023enabling} provided in Fig. \ref{fig:prompt_ans_citation}. The fine-tuned model generates the answer with citations, following the RAG architecture. We first retrieve the 1,000 most relevant PubMed documents for the question using Pyserini from the indexed PubMed corpus. We applied reranker\footnote {cross-encoder/ms-marco-MiniLM-L-6-v2} to re-rank the BM25-retrieved documents. The top-10 re-ranked documents are used to generate the answer by prompting a fine-tuned LLaMA2 7 B model (Fig. \ref{fig:prompt_ans_citation}).

\begin{figure}[t]
\centering
\begin{promptbox}{Answer Generation with Citations  Prompt}
\small
\textbf{Instruction}: Write an accurate, engaging, and concise answer for the given question using only the provided search results (some of which might be irrelevant) and cite them properly. Use an unbiased and journalistic tone. Always cite for any factual claim. When citing several search results, use [1][2][3]. Cite at least one document and at most three documents in each sentence. If multiple documents support the sentence, only cite a minimum sufficient subset of the documents. \\
    \textbf{Documents}: \\
    \{Document 1\} \\
      \{Document 2 \} \\
     ... \\
     \{Document 10\} \\
    \textbf{Question} : \{question\} \\
\end{promptbox}
\caption{Prompt used for answer generation with citations.}
\label{fig:prompt_ans_citation}
\end{figure}
     
\end{enumerate}
The baseline implementations are provided to the participants as a starter kit\footnote{\url{https://github.com/trec-biogen/starter-kit-2025}}.

\section{Assessment}
Due to resource constraints, we conducted assessments of the runs using both expert and automatic evaluation. The details of both evaluations for each task are as follows:

\subsection{Task A (Grounding Answer)} \label{sec:task-a-eval}

\subsubsection{Expert Evaluation}
We constructed a comprehensive citation pool considering 10 selected topics and one top-priority run submitted by each participating team, including the baseline run. The pooling strategy was designed to capture citations that were repeatedly retrieved across multiple runs, while also ensuring representation by including at least one citation from each run and for each answer sentence. This approach was intended to balance coverage and diversity across submitted runs and topics. In total, 244 PubMed abstracts were manually assessed and assigned to one of five categories based on their relationship to the corresponding answer sentence: \textit{supporting}, \textit{partially supporting}, \textit{contradicting}, \textit{neutral}, or \textit{irrelevant}. Using manually assessed PubMed documents, we evaluated the top-priority runs under the Strict and Relaxed evaluation settings. 
Given an answer sentence and its associated set of documents judged as 
$D_{sup}$ (\textbf{supports}), $D_{psup}$ (\textbf{partial supports}), 
and $D_{con}$ (\textbf{contradicts}) and $P_{sup}$ and $P_{con}$ be the subsets of documents that have been manually judged and predicted as supporting and contradicting, respectively.

\begin{itemize}
    \item \textbf{Strict Evaluation.} 
    In this setting, for supports, only predicted documents that are in $D_{sup}$ are considered correct. 
    For the contradict class, only predicted documents in $D_{con}$ are considered correct. 
    Precision is computed as follows:
    \[
    \text{Precision}_{sup}^{strict} = \frac{|P_{sup} \cap D_{sup}|}{|P_{sup}|}, \quad
    \text{Precision}_{con}^{strict} = \frac{|P_{sup}  \cap D_{con}|}{|P_{con}|}
    \]
  Since the pool judgment did not evaluate all model predictions, we use soft recall, where recall is 1 if at least one predicted document is found among the manually judged ground-truth documents.
    \[
    \text{SoftRecall}_{sup}^{strict} =
    \begin{cases}
    1 & \text{if } P_{sup} \cap D_{sup} \neq \emptyset \\
    0 & \text{otherwise}
    \end{cases}, \quad
    \text{SoftRecall}_{con}^{strict} =
    \begin{cases}
    1 & \text{if } P_{con} \cap D_{con} \neq \emptyset \\
    0 & \text{otherwise}
    \end{cases}
    \]

    \item \textbf{Relaxed Evaluation.} 
    In this setting, for supports, predicted documents that are in either $D_{sup}$ or $D_{psup}$ are considered correct. 
    For the contradict class, it remains the same as a strict setting. 
    Precision and recall are computed as follows:
    \[
    \text{Precision}_{sup}^{relaxed} = \frac{|P_{sup}^J \cap (D_{sup} \cup D_{psup})|}{|P_{sup}^J|}, \quad
    \text{Precision}_{con}^{relaxed} = \text{Precision}_{con}^{strict}
    \]
    \[
    \text{SoftRecall}_{sup}^{relaxed} =
    \begin{cases}
    1 & \text{if } P_{sup} \cap (D_{sup} \cup D_{psup}) \neq \emptyset \\
    0 & \text{otherwise}
    \end{cases}, \quad
    \text{SoftRecall}_{con}^{relaxed} = \text{SoftRecall}_{con}^{strict}
    \]
\end{itemize}

After computing precision and recall for each answer sentence, we averaged across topics, and the final result is computed by averaging across all 10 topics considered for pooled judgment. 

\subsubsection{Automatic Evaluation}
For all submitted runs, we used the BioACE \cite{gupta2026bioaceautomatedframeworkbiomedical} evaluation framework for citation evaluations. In particular, we leverage the nugget-based evaluation using LLama-3.3, where, given answer-sentence and cited-document nuggets, the model is prompted to classify the answer-sentence and document pair as \textit{Supports}, \textit{Contradicts}, \textit{Neutral}, or \textit{Not relevant}. We used the nugget-generation approach discussed in \cite{bartels-etal-2025-large}. We treat the generated predictions as ground truth and compute macro-level precision, recall, and F1-score for the support and contradict classes. 

\subsection{Task B (Reference Attribution)} \label{sec:task-b-eval}

\subsubsection{Expert Evaluation}
We follow the BioGen 2024 \cite{biogen2024TREC} evaluation and focused on evaluating the runs on \textbf{(1)} the quality and factuality of the text generated by systems, \textbf{(2)} assessment of the citation against the answer sentence, \textbf{(3)} relevance of the document used by the systems while generating the answers.

\paragraph{Answer quality}

\begin{itemize}
     \item \textbf{Answer Accuracy} (measured at the run level) -- indicates how many answers, out of the total topics, were judged acceptable (i.e., considered to answer the question at least partially) for each run. 
      \begin{equation}
            \text{Accuracy}_{run} = \frac{\text{Number of Acceptable Answers}}{\text{Total Number of Topics}}
        \end{equation}

    \item \textbf{Answer Completeness (Recall)} (answer level) -- measures how many of the answer aspects (pooled across all submitted runs from all participating teams) are covered within a single answer to a question. For the initial evaluation, we cluster the sentences using embeddings from the SentenceTransformer\footnote{sentence-transformers/all-mpnet-base-v2} and SimCSE\footnote{princeton-nlp/sup-simcse-roberta-large} models. Under strict evaluation, only sentences judged as required and supported by evidence are considered. In lenient evaluation, all sentences judged as required are included. In relaxed evaluation, borderline sentences are additionally considered alongside required sentences. The number of aspects for automated clustering is fixed at 10 for K-means. 
    
        \begin{equation}
            \text{Completeness}_{answer} = \frac{\text{Number of Distinct Clusters Containing Sentences from An Answer}}{\text{Number of Clusters}}
        \end{equation}

    \item \textbf{Answer Precision} (answer level) -- measures how many assertions in an answer were judged as required or acceptable. The strict, lenient, and relaxed evaluation settings apply here as well. 

          \begin{equation}
            \text{Precision}_{answer} = \frac{\text{Number of Generated Required Answer Sentences}}{\text{Total Number of Generated Answer Sentences}}
        \end{equation}

    \item \textbf{Redundancy Score} (answer level): penalizes a system for producing \textit{unnecessary} answer sentences. It reflects the informativeness of the generated answers. 
        \begin{equation}
            \text{Redundancy Score} =  \frac{\text{Number of Generated Unnecessary Answer Sentences}}{\text{Total Number of Generated Answer Sentences}}
        \end{equation}
 
    \item \textbf{Irrelevant Score} (answer level): penalizes a system for generating \textit{inappropriate}/\textit{potentially harmful} answer sentences. %It measures how misleading or harmful the generated answers are. 
        \begin{equation}
            \text{Harmfulness Score} =  \frac{\text{Number of Generated Answer Harmful Sentences}}{\text{Total Number of Generated Answer Sentences}}
        \end{equation}
\end{itemize}

\paragraph{Citation Quality}

Answer statements may be supported or contradicted by the documents cited as references. Some sentences may contain no references, or may include references that are only topically related or not relevant. The following metric measures reference quality.

\begin{itemize}
  \item \textbf{Citation Coverage}: measures how well the required and borderline generated answer sentences are backed by the appropriate (judged as supports) citations. 
      \begin{equation}
      \label{eq:cc}
      \footnotesize
          \text{Citation Coverage} =  \frac{\text{Number of Systems Generated Answer Sentences with One or More Supportive Citation}}{\text{Total Number of Generated Answer Sentences}}
       \end{equation}
         \item \textbf{Citation Support Rate}: assesses how well the system-predicted citations are aligned with the human-judged support citations.
    \begin{equation}
    \label{eq:csr}
            \text{Citation Support Rate} =  \frac{\text{Number of \textit{Supports} Citations}}{\text{Total Number of Citations}}
        \end{equation}
          \item \textbf{Citation Contradict Rate}: penalizes the answers that are providing documents assessed as \textit{Contradicting} the statement. Note that in a fact-verification task, this measure may show how well a system is finding contradictory evidence. 
     \begin{equation}
       \label{eq:ccr}
            \text{Citation Contradict Rate} =  \frac{\text{Number of \textit{Contradict} Citations}}{\text{Total Number of Citations}}
       \end{equation}

\end{itemize}
\paragraph{Document relevancy}
By pooling all documents judged relevant for a given topic, we can compute standard recall and precision. Relevant documents include those judged as supporting, contradicting, or neutral.

\begin{itemize}
\item Recall:
    \begin{equation}
            \text{Recall} = \frac{\text{Number of relevant retrieved documents}}{\text{all relevant documents}}
        \end{equation}

        \item Precision:
    \begin{equation}
            \text{Precision} = \frac{\text{Number of relevant retrieved documents}}{\text{Number of references provided}}
        \end{equation}

   \end{itemize}
   
Each of the aforementioned metrics was computed per topic and then averaged across all topics in the test collection to obtain the final scores.

\subsubsection{Automatic Evaluation}
Following Task A, we used the BioACE \cite{gupta2026bioaceautomatedframeworkbiomedical} evaluation framework to evaluate answers. The evaluation measures \textit{Completeness}, \textit{Correctness},  \textit{Nugget Precision}, and \textit{Nugget Recall} against the ground-truth answer/nuggets. The \textit{completeness} measures the portion of the generated answer that the model considered \textit{Required} to be part of the answer.  The \textit{correctness} measures the portion of the generated answer that aligns with the relevant documents for the topic. In the nugget-based evaluation, the nugget is first generated for the given answer, and then aligned with the ground-truth nugget using the relaxed matching scheme discussed in BioACE. Thereafter, precision and recall are computed. As with the expert evaluation, these metrics are computed per topic and then averaged across all topics in the test collection to produce the final scores. 
We use the BioACE evaluation framework for the citation evaluation. In particular, we use LLama-3.3, where, given an answer-sentence and a cited-document, the model is prompted to classify the answer-sentence and document pair as \textit{Supports}, \textit{Contradicts}, or \textit{Neutral}. Given the predicted label, we compute the citation coverage, citation support rate, and citation contradiction rate using Eq. \ref{eq:cc}, \ref{eq:csr}, and \ref{eq:ccr}, respectively. 

\begin{figure}
    \centering
    \includegraphics[width=\linewidth]{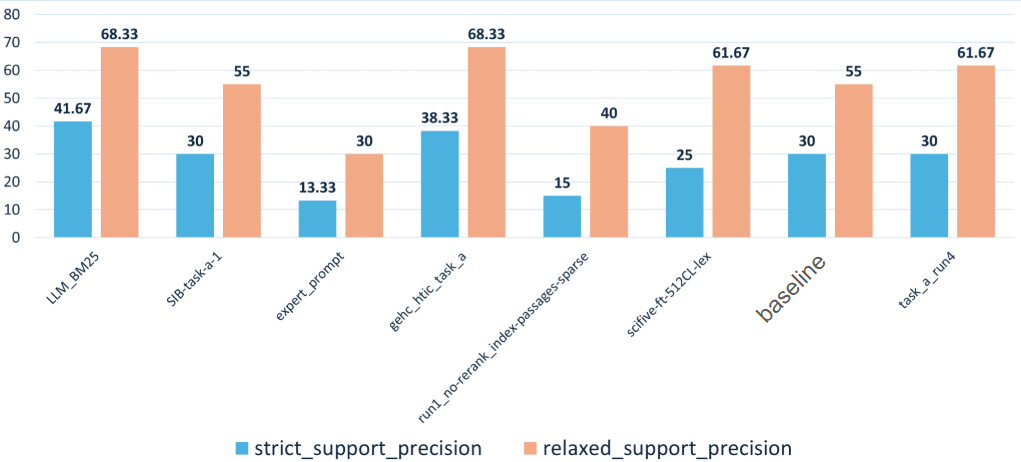}
    \caption{\textbf{(Task A)} Comparison of the submitted runs using the expert evaluation scheme in terms of Precision (Strict and Relaxed) for the \textit{Supports} class.}
    \label{fig:task-a-supports-precision}
\end{figure}
\begin{figure}
    \centering
    \includegraphics[width=\linewidth]{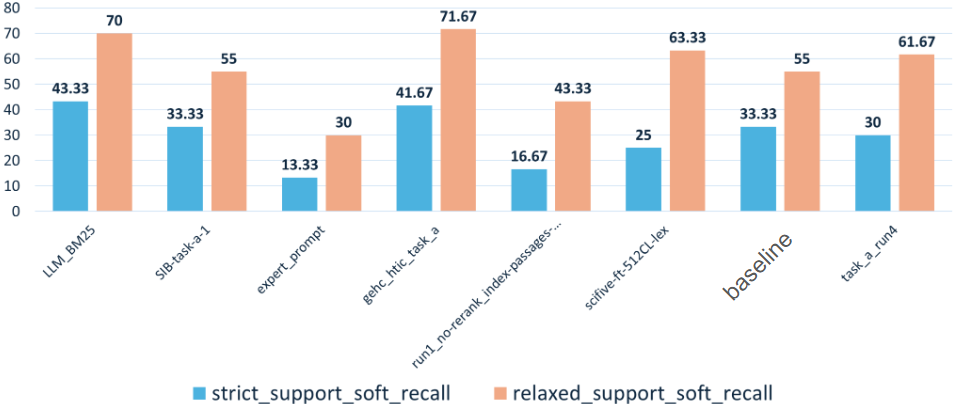}
    \caption{\textbf{(Task A)} Comparison of the submitted runs using the expert evaluation scheme in terms of SoftRecall (Strict and Relaxed) for the \textit{Supports} class.}
    \label{fig:task-a-supports-recall}
\end{figure}
\begin{figure}
    \centering
    \includegraphics[width=\linewidth]{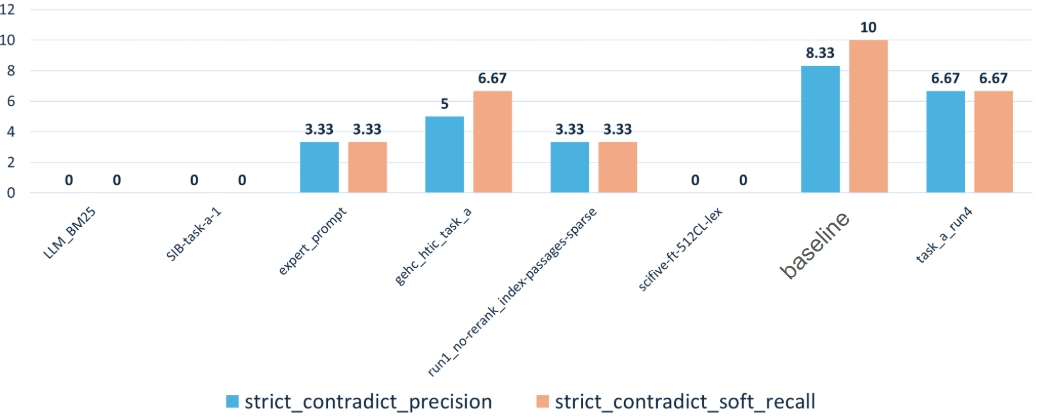}
    \caption{\textbf{(Task A)} Comparison of the submitted runs using the expert evaluation scheme in terms of Precision and SoftRecall for the \textit{Contradicts} class. Please note that for the \textit{Contradicts} class, strict and relaxed results remain the same.}
    \label{fig:task-a-contadicts-precision-recall}
\end{figure}

% \usepackage{graphicx}
\begin{table}[]
\centering
\resizebox{\columnwidth}{!}{%
\begin{tabular}{l|l|lll|lll}
\hline
\multirow{2}{*}{\textbf{Team Name}} &
  \multirow{2}{*}{\textbf{Run Name/File Name}} &
  \multicolumn{3}{c|}{\textbf{Supports}} &
  \multicolumn{3}{c}{\textbf{Contradicts}} \\ \cline{3-8} 
 &
   &
  {\textbf{Precision}} &
  {\textbf{Recall}} &
  \textbf{F1-Score} &
  {\textbf{Precision}} &
  {\textbf{Recall}} &
  \textbf{F1-Score} \\ \hline \hline
\multirow{5}{*}{InfoLab} &
  task\_a\_run1 &
  {39.95} &
  {47.59} &
  42.07 &
  {7.99} &
  {13.92} &
  9.73 \\ 
 &
  task\_a\_run2 &
  {52.75} &
  {56.8} &
  53.41 &
  {14.09} &
  {19.42} &
  15.67 \\ 
 &
  task\_a\_run4 &
  {52.92} &
  {60.3} &
  54.49 &
  {10.74} &
  {15.12} &
  11.85 \\ 
 &
  task\_a\_run6\_A &
  {66.92} &
  {71.17} &
  67.23 &
  {12.71} &
  {17.65} &
  14.15 \\ 
 &
  task\_a\_run3 &
  {18.56} &
  {18.56} &
  18.56 &
  {0.52} &
  {0.52} &
  0.52 \\ \hline
\multirow{7}{*}{SIB} &
  SIB-task-a-3 &
  {15.12} &
  {26.8} &
  18.42 &
  {0} &
  {0} &
  0 \\ 
 &
  SIB-task-a-6 &
  {36.25} &
  {42.14} &
  37.82 &
  {1.98} &
  {2.58} &
  2.15 \\ 
 &
  SIB-task-a-2 &
  {47.85} &
  {64.09} &
  52.3 &
  {3.69} &
  {5.67} &
  4.21 \\ 
 &
  SIB-task-a-5 &
  {46.22} &
  {64} &
  51.17 &
  {2.06} &
  {3.95} &
  2.56 \\ 
 &
  SIB-task-a-4 &
  {17.01} &
  {32.47} &
  21.31 &
  {0} &
  {0} &
  0 \\ 
 &
  SIB-task-a-1 &
  {52.41} &
  {74.23} &
  58.87 &
  {0} &
  {0} &
  0 \\ 
&
  SIB-task-a-7 &
  {39.52} &
  {45.02} &
  41.08 &
  {2.06} &
  {2.58} &
  2.23 \\ \hline
Baseline &
  TEST &
  {51.03} &
  {44.07} &
  44.34 &
  {3.44} &
  {8.08} &
  4.67 \\ \hline
polito &
  scifive-ft-512CL-lex &
  {52.58} &
  {64.54} &
  55.81 &
  {4.04} &
  {6.7} &
  4.79 \\ \hline
\multirow{2}{*}{dal} &
  emotional\_prompt &
  {50.6} &
  {67.23} &
  55.53 &
  {1.29} &
  {1.29} &
  1.2 \\ 
 &
  expert\_prompt &
  {45.62} &
  {55.67} &
  48.8 &
  {0.52} &
  {0.26} &
  0.34 \\ \hline
\multirow{2}{*}{CLaC} &
  LLM\_NLI\_BM25 &
  {67.18} &
  {74.36} &
  67.74 &
  {3.61} &
  {7.73} &
  4.57 \\ 
 &
  LLM\_BM25 &
  {66.75} &
  {67.46} &
  64.1 &
  {3.95} &
  {7.6} &
  4.77 \\ \hline
\multirow{4}{*}{uniud} &
  run4\_rerank\_index-passages-dense &
  {16.58} &
  {8.79} &
  10.52 &
  {2.58} &
  {7.22} &
  3.76 \\ 
 &
  run1\_no-rerank\_index-passages-sparse &
  {40.89} &
  {41.78} &
  39.1 &
  {1.55} &
  {2.66} &
  1.87 \\ 
 &
  run3\_no-rerank\_index-passages-dense &
  {26.55} &
  {19} &
  19.82 &
  {3.61} &
  {9.19} &
  5.07 \\ 
 &
  run2\_rerank\_index-passages-sparse &
  {39.18} &
  {44.86} &
  38.85 &
  {2.23} &
  {1.98} &
  2.03 \\ \hline
GEHC-HTIC &
  gehc\_htic\_task\_a &
  {56.7} &
  {57.37} &
  53.53 &
  {6.62} &
  {14} &
  8.57 \\ \hline \hline
\end{tabular}%
}
\caption{\textbf{(Task A)} Performance comparison of the submitted runs using the automated evaluation scheme in terms of precision, recall, and F1-score for Supports and Contradicts classes. }
\label{tab:task-a-automatic-eval}
\end{table}

\begin{table}[]
\resizebox{\columnwidth}{!}{%
\begin{tabular}{l|l|l}
\hline
\textbf{Team   Name}           & \textbf{Run Name}                                                       & \textbf{Accuracy} \\ \hline
\hline
\multirow{2}{*}{dal}           
& rrf\_monot5-msmarco\_llama70b                                           & 90                \\ 
& rrf\_monot5-msmarco\_deepseek-r1                                        & 100               \\ \hline

\multirow{2}{*}{hltcoe-multiagt} 
& hltcoe-multiagt.llama70B.lg-w-ret\_p-nt-s\_cite - [X].jsonl     & 100               \\ 
& hltcoe-multiagt.llama70B.lg-w-ret\_lpq-nt-s\_cite - [X].jsonl   & 100               \\ \hline

GEHC                           
& task\_b\_output\_run\_gehc - [X].json                   & 80                \\ \hline

\multirow{3}{*}{uniud}           
& run3\_no\_rerank\_index-passages-dense\_Llama-3.1-8B-Instruct           & 93.33             \\ 
& run1\_no-rerank\_index-passages-sparse\_Llama-3.1-8B-Instruct           & 96.67             \\ 
& run2\_rerank\_index-passages-sparse\_Llama-3.1-8B-Instruct              & 93.33             \\ \hline

\multirow{2}{*}{UAmsterdam}      
& UAmsterdam\_bergen\_llama-8b - [X].jsonl                         & 100               \\ 
& UAmsterdam\_bergen\_llama-70b - [X].jsonl                        & 100               \\ \hline

EvalHLTCOE                     
& hltcoe-eval-svc-smoothed-sonnet - [X].jsonl                     & 100               \\ \hline

\multirow{4}{*}{hltcoe-rerank} 
& hltbio-lg.searcher                                                      & 93.33             \\ 
& hltcoe-rerank.llama70B.lg-w-ret\_plq-nt-s\_cite - [X].jsonl    & 100               \\ 
& hltbio-gpt5.searcher                                                    & 100               \\ 
& hltbio-lg.crux                                                          & 93.33             \\ \hline

Baseline                       
& task\_b\_baseline\_output.json                                          & 86.67             \\ \hline

UDInfo                         
& task\_b\_output\_reranker\_sum - [X].json                   & 100               \\ \hline
\hline
\end{tabular}%
}
\caption{\textbf{(Task B)} Performance comparison of the submitted runs for the answer quality using the expert evaluation scheme in terms of \textit{Accuracy} metric.}
\label{tab:res-acc}
\end{table}

\begin{table}[t]
\centering
\resizebox{\columnwidth}{!}{%
\begin{tabular}{l|l|ccc}
\hline
\textbf{Team Name} & \textbf{Run Name} & \textbf{Precision} & \textbf{Redundancy} & \textbf{Harmfulness} \\
\hline \hline

\multirow{2}{*}{dal}
& rrf\_monot5-msmarco\_llama70b & 65.16 & 12.2 & 3.18 \\
& rrf\_monot5-msmarco\_deepseek-r1 & 68.05 & 14.96 & 0 \\
\hline

\multirow{2}{*}{hltcoe-multiagt}
& hltcoe-multiagt.llama70B.lg-w-ret\_p-nt-s\_cite - [X].jsonl & 63.79 & 24.25 & 1.3 \\
& hltcoe-multiagt.llama70B.lg-w-ret\_lpq-nt-s\_cite - [X].jsonl & 54.52 & 27.75 & 4.24 \\
\hline

\multirow{3}{*}{uniud}
& run3\_no\_rerank\_index-passages-dense\_Llama-3.1-8B-Instruct & 69.49 & 16.78 & 0 \\
& run1\_no-rerank\_index-passages-sparse\_Llama-3.1-8B-Instruct & 61.3 & 16.56 & 1.11 \\
& run2\_rerank\_index-passages-sparse\_Llama-3.1-8B-Instruct & 50.44 & 25.33 & 1.37 \\
\hline

\multirow{2}{*}{UAmsterdam}
& UAmsterdam\_bergen\_llama-8b - [X].jsonl & 60.35 & 19.8 & 3.1 \\
& UAmsterdam\_bergen\_llama-70b - [X].jsonl & 71.55 & 14.12 & 0.67 \\
\hline

EvalHLTCOE
& hltcoe-eval-svc-smoothed-sonnet - [X].jsonl & 57.61 & 24.32 & 1.43 \\
\hline

\multirow{4}{*}{hltcoe-rerank}
& hltbio-lg.searcher & 46.78 & 17.38 & 2.59 \\
& hltcoe-rerank.llama70B.lg-w-ret\_plq-nt-s\_cite - [X].jsonl & 53.29 & 26.53 & 2.38 \\
& hltbio-gpt5.searcher & 46.22 & 26.15 & 1.93 \\
& hltbio-lg.crux & 68.59 & 12.66 & 3.29 \\
\hline

Baseline
& task\_b\_baseline\_output.json & 52.66 & 19.77 & 1.44 \\
\hline

UDInfo
& task\_b\_output\_reranker\_sum - [X].json & 49.15 & 35.94 & 3.01 \\
\hline

GEHC
& task\_b\_output\_run\_gehc - [X].json & 44.93 & 20.92 & 0.74 \\
\hline \hline

\end{tabular}%
}
\caption{\textbf{(Task B)} Evaluation of submitted runs for answer quality using the expert evaluation scheme in terms of \textit{Precision}, \textit{Redundancy}, and \textit{Harmfulness}.}
\label{tab:res-quality}
\end{table}

% \usepackage{graphicx}

\begin{table}[t]
\centering
\resizebox{\columnwidth}{!}{%
\begin{tabular}{l|l|ccc}
\hline
\textbf{Team Name} &
\textbf{Run Name} &
\textbf{Recall (S+R)} &
\textbf{Recall (R)} &
\textbf{Recall (R+B)} \\
\hline \hline

\multirow{2}{*}{dal}
& rrf\_monot5-msmarco\_llama70b & 23 & 22 & 24.33 \\
& rrf\_monot5-msmarco\_deepseek-r1 & 34.67 & 38 & 40.33 \\
\hline

\multirow{2}{*}{hltcoe-multiagt}
& hltcoe-multiagt.llama70B.lg-w-ret\_p-nt-s\_cite - [X].jsonl & 37.67 & 41.67 & 44.33 \\
& hltcoe-multiagt.llama70B.lg-w-ret\_lpq-nt-s\_cite - [X].jsonl & 34.67 & 39.67 & 46 \\
\hline

\multirow{3}{*}{uniud}
& run3\_no\_rerank\_index-passages-dense\_Llama-3.1-8B-Instruct & 14 & 33.33 & 34.33 \\
& run1\_no-rerank\_index-passages-sparse\_Llama-3.1-8B-Instruct & 8.33 & 29.33 & 30.33 \\
& run2\_rerank\_index-passages-sparse\_Llama-3.1-8B-Instruct & 10.67 & 26.33 & 28 \\
\hline

\multirow{2}{*}{UAmsterdam}
& UAmsterdam\_bergen\_llama-8b - [X].jsonl & 29 & 37 & 43 \\
& UAmsterdam\_bergen\_llama-70b - [X].jsonl & 28 & 33.33 & 34.67 \\
\hline

EvalHLTCOE
& hltcoe-eval-svc-smoothed-sonnet - [X].jsonl & 21.33 & 28.67 & 32.33 \\
\hline

\multirow{4}{*}{hltcoe-rerank}
& hltbio-lg.searcher & 24.67 & 28.33 & 34.33 \\
& hltcoe-rerank.llama70B.lg-w-ret\_plq-nt-s\_cite - [X].jsonl & 28.67 & 35.33 & 43 \\
& hltbio-gpt5.searcher & 37.67 & 45 & 54 \\
& hltbio-lg.crux & 28.33 & 30 & 35.33 \\
\hline

Baseline
& task\_b\_baseline\_output.json & 14 & 22.33 & 25.67 \\
\hline

UDInfo
& task\_b\_output\_reranker\_sum - [X].json & 24.33 & 28.67 & 33 \\
\hline

GEHC
& task\_b\_output\_run\_gehc - [X].json & 21.33 & 31 & 36.33 \\
\hline \hline

\end{tabular}%
}
\caption{\textbf{(Task B)} Comparison of submitted runs based on answer quality using the expert evaluation scheme in terms of the Recall metric. The abbreviations are as follows: (\textbf{S+R}) only answer sentences judged required and supported by evidence were clustered; (\textbf{R}) answer sentences judged required were clustered; and (\textbf{R+B}) answer sentences judged required or borderline were clustered. The answer sentence representations are obtained using the Sentence Transformers.}
\label{tab:res-recall}
\end{table}

\begin{table}[t]
\centering
\resizebox{\columnwidth}{!}{%
\begin{tabular}{l|l|ccc}
\hline
\textbf{Team Name} &
\textbf{Run Name} &
\begin{tabular}[c]{@{}l@{}}\textbf{Citation}\\ \textbf{Coverage}\end{tabular} &
\begin{tabular}[c]{@{}c@{}}\textbf{Citation Support}\\ \textbf{Rate}\end{tabular} &
\begin{tabular}[c]{@{}c@{}}\textbf{Citation Contradict}\\ \textbf{Rate}\end{tabular}
\\
\hline \hline

\multirow{2}{*}{dal}
& rrf\_monot5-msmarco\_llama70b & 78.74 & 73.79 & 3.88 \\
& rrf\_monot5-msmarco\_deepseek-r1 & 85.24 & 85.53 & 4.4 \\
\hline

\multirow{2}{*}{hltcoe-multiagt}
& hltcoe-multiagt.llama70B.lg-w-ret\_p-nt-s\_cite - [X].jsonl & 84.31 & 82.52 & 1.97 \\
& hltcoe-multiagt.llama70B.lg-w-ret\_lpq-nt-s\_cite - [X].jsonl & 74.48 & 73.02 & 3.02 \\
\hline

\multirow{3}{*}{uniud}
& run3\_no\_rerank\_index-passages-dense\_Llama-3.1-8B-Instruct & 30.79 & 22.64 & 3.17 \\
& run1\_no-rerank\_index-passages-sparse\_Llama-3.1-8B-Instruct & 32.88 & 29.38 & 1.09 \\
& run2\_rerank\_index-passages-sparse\_Llama-3.1-8B-Instruct & 46.59 & 35.86 & 1.58 \\
\hline

\multirow{2}{*}{UAmsterdam}
& UAmsterdam\_bergen\_llama-8b - [X].jsonl & 73.83 & 92.38 & 3.49 \\
& UAmsterdam\_bergen\_llama-70b - [X].jsonl & 87.62 & 79.03 & 2.04 \\
\hline

EvalHLTCOE
& hltcoe-eval-svc-smoothed-sonnet - [X].jsonl & 63.98 & 56.79 & 3.34 \\
\hline

\multirow{4}{*}{hltcoe-rerank}
& hltbio-lg.searcher & 66.29 & 63.76 & 6.53 \\
& hltcoe-rerank.llama70B.lg-w-ret\_plq-nt-s\_cite - [X].jsonl & 70.39 & 66.63 & 5.35 \\
& hltbio-gpt5.searcher & 74.69 & 69.99 & 5.11 \\
& hltbio-lg.crux & 76.13 & 73.46 & 1.93 \\
\hline

Baseline
& task\_b\_baseline\_output.json & 38.36 & 44.99 & 6.46 \\
\hline

UDInfo
& task\_b\_output\_reranker\_sum - [X].json & 53.08 & 69.07 & 1.38 \\
\hline

GEHC
& task\_b\_output\_run\_gehc - [X].json & 50.11 & 51.72 & 1.29 \\
\hline \hline

\end{tabular}%
}
\caption{\textbf{(Task B)} Comparison of submitted runs based on citation quality under the expert evaluation scheme using multiple metrics.}
\label{tab:res-citation-quality}
\end{table}

% \usepackage{graphicx}

\begin{table}[t]
\centering
\resizebox{\columnwidth}{!}{%
\begin{tabular}{l|l|cc}
\hline
\textbf{Team Name} & \textbf{Run Name} & \textbf{Recall} & \textbf{Precision} \\
\hline

\multirow{2}{*}{dal}
& rrf\_monot5-msmarco\_llama70b & 4.27 & 86.19 \\
& rrf\_monot5-msmarco\_deepseek-r1 & 5.83 & 97.22 \\
\hline

\multirow{2}{*}{hltcoe-multiagt}
& hltcoe-multiagt.llama70B.lg-w-ret\_p-nt-s\_cite - [X].jsonl & 8.63 & 95.04 \\
& hltcoe-multiagt.llama70B.lg-w-ret\_lpq-nt-s\_cite - [X].jsonl & 9.16 & 94.23 \\
\hline

\multirow{3}{*}{uniud}
& run3\_no\_rerank\_index-passages-dense\_Llama-3.1-8B-Instruct & 3.29 & 50.13 \\
& run1\_no-rerank\_index-passages-sparse\_Llama-3.1-8B-Instruct & 4.18 & 73.52 \\
& run2\_rerank\_index-passages-sparse\_Llama-3.1-8B-Instruct & 4.65 & 75.62 \\
\hline

\multirow{2}{*}{UAmsterdam}
& UAmsterdam\_bergen\_llama-8b - [X].jsonl & 5.35 & 97.36 \\
& UAmsterdam\_bergen\_llama-70b - [X].jsonl & 5.65 & 93.43 \\
\hline

EvalHLTCOE
& hltcoe-eval-svc-smoothed-sonnet - [X].jsonl & 6.9 & 84.26 \\
\hline

\multirow{4}{*}{hltcoe-rerank}
& hltbio-lg.searcher & 7.45 & 86.48 \\
& hltcoe-rerank.llama70B.lg-w-ret\_plq-nt-s\_cite - [X].jsonl & 7.6 & 90.57 \\
& hltbio-gpt5.searcher & 13.57 & 94.83 \\
& hltbio-lg.crux & 5.16 & 88.26 \\
\hline

Baseline
& task\_b\_baseline\_output.json & 2.26 & 67.83 \\
\hline

UDInfo
& task\_b\_output\_reranker\_sum - [X].json & 6.15 & 88.46 \\
\hline

GEHC
& task\_b\_output\_run\_gehc - [X].json & 3.97 & 74 \\
\hline \hline

\end{tabular}%
}
\caption{\textbf{(Task B)} Evaluation of submitted runs for document relevancy using the expert evaluation scheme focusing on \textit{Precision} and \textit{Recall} metrics.}
\label{tab:res-doc-relevance}
\end{table}

% \usepackage{multirow}
% \usepackage{graphicx}
\begin{table}[]
\centering
\resizebox{\columnwidth}{!}{%
\begin{tabular}{l|l|cccc||ccc}
\hline
\textbf{Team Name} &
  \textbf{Run Name} &
\textbf{\begin{tabular}[c]{@{}l@{}}Nugget\\ Precision\end{tabular}} &
\textbf{\begin{tabular}[c]{@{}l@{}}Nugget\\ Recall\end{tabular}} &
  \textbf{Completeness} &
  \textbf{Correctness} &
  \textbf{\begin{tabular}[c]{@{}l@{}}Citation\\ Coverage\end{tabular}} &
  \textbf{\begin{tabular}[c]{@{}c@{}}Citation Support\\ Rate\end{tabular}} &
  \textbf{\begin{tabular}[c]{@{}c@{}}Citation Contradict\\ Rate\end{tabular}} \\ \hline \hline
\multirow{4}{*}{CLaC}            & MedhopQA\_FAISS                                    & 86.3  & 37.56 & 78.06 & 67.45 & 93.3  & 91.99 & 0.59  \\ 
                                 & simpleQA\_BM25                                     & 91.23 & 35.59 & 75.97 & 66.42 & 91.92 & 80    & 1.79  \\ 
                                 & MedHopQA\_BM25                                     & 87.91 & 37.02 & 76.07 & 68.39 & 93.21 & 93.88 & 0.19  \\ 
                                 & simpleQA\_Hybrid                                   & 92.15 & 35.16 & 82.67 & 67.5  & 96.1  & 82.81 & 2.26  \\ \hline
\multirow{5}{*}{CLaC\_Lab}       & System\_A\_Baseline\_Plus\_Qwen                    & 88.31 & 32.34 & 77.63 & 66.9  & 74.87 & 97.38 & 0.44  \\ 
                                 & System\_D\_medcpt\_refined                         & 93.83 & 36.73 & 72.56 & 67.33 & 92.21 & 87.46 & 1.39  \\ 
                                 & System\_C\_stage3\_MedCPT\_bi\_cross               & 92.28 & 36.41 & 69.53 & 65.74 & 94.9  & 90.37 & 0.46  \\ 
                                 & System\_B\_stage3\_TFIDF                           & 89.04 & 35.15 & 67.7  & 64.19 & 95.45 & 91.12 & 0.39  \\ 
                                 & System\_E\_wide\_stage3\_TFIDF                     & 90.99 & 35.8  & 66.77 & 64.56 & 95.83 & 87.32 & 0.72  \\ \hline
{dal}             & rrf\_mistral-24\_monot5-msmarco\_deepseek-r1       & 91.95 & 35.83 & 79.78 & 67.19 & 96.02 & 97.95 & 0.41  \\ \hline
                                 % & rrf\_monot5-msmarco\_deepseek-r1\_no-val           & 93.9  & 35.94 & 80.28 & 63.07 & 97.73 & 97.17 & 0     \\ 
                                 % & emotional\_prompt\_b\_deepseek                     & 88.23 & 36.25 & 80.95 & 68.42 & 94.58 & 82.78 & 0.62  \\ 
                                 % & expert\_prompt\_b\_deepseek                        & 91.19 & 34.4  & 79.78 & 67.92 & 85.71 & 80.99 & 0.87  \\ 
                                 % & rrf\_monot5-msmarco\_llama70b\_no-val              & 94.68 & 37.86 & 89.99 & 67.26 & 99.3  & 97.75 & 1.12  \\ 
                                 % & agent\_faiss\_monot5-msmarco\_deepseek-r1          & 94.96 & 38.42 & 82.94 & 64.32 & 85.9  & 98.2  & 0.45  \\ 
                                 % & rrf\_monot5-msmarco\_llama70b                      & 93.95 & 37.02 & 86.29 & 66.7  & 99.36 & 98.33 & 1.11  \\ 
                                 % & rrf\_mistral-24\_monot5-msmarco\_deepseek-r1       & 89.42 & 36.97 & 86.7  & 64.86 & 94.74 & 93.89 & 1.31  \\ 
                                 % & agent\_faiss\_monot5-msmarco\_deepseek-r1\_no\_val & 94.42 & 38.12 & 82.5  & 64.96 & 88.63 & 93.87 & 0.77  \\ \hline
GEHC-HTIC                        & GEHC-HTIC\_pubmedbert\_medcpt\_gpt\_4o             & 89.02 & 35.42 & 59.83 & 67.56 & 98.77 & 95.39 & 0     \\ \hline
\multirow{4}{*}{h2oloo}          & h2oloo\_rr\_g41\_t50                               & 84.29 & 32.5  & 89.17 & 60.5  & 98.59 & 93.82 & 0.29  \\ 
                                 & h2oloo\_rr\_q3-30b\_nc                             & 88.41 & 37.24 & 84.48 & 67.16 & 0     & 0     & 0     \\ 
                                 & h2oloo\_rr\_q3-30b\_t50                            & 90.25 & 35.47 & 78.78 & 66.45 & 87.69 & 90.91 & 1.3   \\ 
                                 & h2oloo\_rr\_q3-30b\_t20                            & 87.99 & 35.16 & 80.67 & 63.75 & 88.54 & 88.22 & 0.64  \\ \hline
% \multirow{8}{*}{hltbio-retrieve} & hltbio-lg.qwen                                     & 90.97 & 36.86 & 68.8  & 68.33 & 90.41 & 89.31 & 1.05  \\ 
%                                  & hltbio-lg.jina                                     & 89.89 & 36.91 & 70.49 & 68.53 & 91.59 & 96.73 & 0     \\ 
%                                  & hltbio-lg.fsrrfprf                                 & 92.52 & 37.82 & 63.77 & 69.67 & 95.77 & 94.52 & 0.52  \\ 
%                                  & hltbio-lg.listllama                                & 92.28 & 37.95 & 69.31 & 68.1  & 96.95 & 94.84 & 0.49  \\ 
%                                  & hltbio-gpt5.searcher                               & 88.13 & 35.61 & 73.03 & 66.18 & 96.97 & 93.52 & 0.97  \\ 
%                                  & hltbio-lg.jina.qwen                                & 93.4  & 39.13 & 71.86 & 69.11 & 93.13 & 96.4  & 0.24  \\ 
%                                  & hltbio-lg.searcher                                 & 89.87 & 37.47 & 72.07 & 68.8  & 94.07 & 94.01 & 0.46  \\ 
%                                  & hltbio-lg.fsrrf                                    & 93.22 & 40.27 & 72.95 & 68.04 & 96.89 & 93.59 & 0.48  \\ \hline
% hltcoe                           & hltbio-lg.crux                                     & 90.47 & 39.07 & 73.84 & 67.93 & 94.69 & 94.05 & 1.12  \\ \hline
\multirow{4}{*}{SIB}             & SIB-task-b-3                                       & 85.74 & 33.81 & 38.98 & 66.32 & 98.21 & 95.39 & 0.66  \\ 
                                 & SIB-task-b-2                                       & 90.29 & 33.28 & 54.3  & 64.19 & 44.71 & 77.78 & 1.23  \\ 
                                 & SIB-task-b-4                                       & 85.1  & 30.52 & 72    & 61.67 & 78.08 & 86.9  & 1.19  \\ 
                                 & SIB-task-b-1                                       & 92.32 & 35.8  & 28.79 & 63.93 & 48.39 & 66.83 & 2.48  \\ \hline
% \multirow{10}{*}{SMDC\_Uniud}    & 1                                                  & 91.04 & 36.77 & 57.28 & 66.19 & 75.13 & 76.4  & 0.75  \\ 
%                                  & 3                                                  & 87.08 & 37.09 & 70.11 & 68.33 & 45.93 & 39.24 & 6.01  \\ 
%                                  & 9                                                  & 90.04 & 36.87 & 71.63 & 66.87 & 69.15 & 79.45 & 0.4   \\ 
%                                  & 4                                                  & 84.42 & 35.7  & 69.21 & 68.01 & 53.47 & 48.61 & 4.33  \\ 
%                                  & 6                                                  & 91.21 & 34.04 & 53.79 & 67.92 & 85.8  & 72.37 & 0.3   \\ 
%                                  & 7                                                  & 81.68 & 29.35 & 42.81 & 69.17 & 70.73 & 47.82 & 8.11  \\ 
%                                  & 11                                                 & 88.77 & 35.95 & 67.7  & 68.24 & 33.16 & 58.08 & 6.57  \\ 
%                                  & 8                                                  & 81.04 & 28.61 & 37.46 & 64.84 & 62.42 & 47.44 & 10.04 \\ 
%                                  & 5                                                  & 90.25 & 35.29 & 55.3  & 70.16 & 93.3  & 79.24 & 0.88  \\ 
%                                  & 2                                                  & 89.4  & 36.28 & 55.55 & 67.62 & 79.5  & 76.35 & 2.03  \\ \hline
\multirow{4}{*}{InfoLab}          & UDInfo\_run4                                       & 89.67 & 38.6  & 82.62 & 66.71 & 64.57 & 85.53 & 0.85  \\ 
                                 & UDInfo\_run5                                       & 89.28 & 34.55 & 50.26 & 64.8  & 93.78 & 89.95 & 1.51  \\ 
                                 & UDInfo\_run1                                       & 87.01 & 34.02 & 70.19 & 65.78 & 77.45 & 89.4  & 0.27  \\ 
                                 & UDInfo\_run4\_B                                    & 91.04 & 36.35 & 74.56 & 66.2  & 76.06 & 95.05 & 0.71  \\ \hline \hline
\end{tabular}%
}
\caption{\textbf{(Task B)} Evaluation of runs in \textit{final submissions} for answer and citation evaluation using the automated evaluation scheme.}
\label{tab:task-b-automated-answer-eval-early}
\end{table}

\begin{table}[]
\centering
\resizebox{\columnwidth}{!}{%
\begin{tabular}{l|l|cccc||ccc}
\hline
\textbf{Team Name} &
  \textbf{Run Name} &
\textbf{\begin{tabular}[c]{@{}l@{}}Nugget\\ Precision\end{tabular}} &
\textbf{\begin{tabular}[c]{@{}l@{}}Nugget\\ Recall\end{tabular}} &
  \textbf{Completeness} &
  \textbf{Correctness} &
  \textbf{\begin{tabular}[c]{@{}l@{}}Citation\\ Coverage\end{tabular}} &
  \textbf{\begin{tabular}[c]{@{}c@{}}Citation Support\\ Rate\end{tabular}} &
  \textbf{\begin{tabular}[c]{@{}c@{}}Citation Contradict\\ Rate\end{tabular}} \\ \hline \hline
Baseline                        & task\_b\_baseline\_output.json.json                                        & 82.23 & 32.5  & 49.73 & 60    & 55.76 & 77.84 & 3.59 \\ \hline
\multirow{10}{*}{hltcoe-rerank} & hltbio-lg-fsrrf.json                                                       & 91.72 & 40.38 & 74.27 & 68.04 & 97.52 & 95.37 & 0.51 \\ 
                                & hltbio-lg.jina.qwen.json                                                   & 93.94 & 39.31 & 71.81 & 69.11 & 93.43 & 97.32 & 0.54 \\ 
                                & hltbio-lg.searcher.json                                                    & 91.22 & 37.87 & 74.04 & 68.8  & 93.47 & 94.12 & 0.51 \\ 
                                & hltbio-lg.listllama.json                                                   & 90.26 & 37.51 & 67.96 & 68.1  & 96.65 & 95.91 & 0.54 \\ 
                                & hltcoe-rerank.llama70B.lg-w-ret\_plq-nt-s\_cite - [X].jsonl.json   & 92.27 & 37.97 & 73.37 & 68.95 & 96.01 & 95.32 & 0.7  \\ 
                                & hltbio-lg.crux.json                                                        & 90.47 & 39.19 & 75.5  & 67.93 & 93.72 & 93.75 & 1.25 \\ 
                                & hltbio-lg.jina.json                                                        & 87.8  & 35.46 & 67.93 & 68.53 & 91.88 & 96.7  & 0.27 \\ 
                                & hltbio-gpt5.searcher.json                                                  & 90.44 & 36.98 & 72.73 & 66.18 & 96.97 & 94.49 & 0.86 \\ 
                                & hltbio-lg.qwen.json                                                        & 90.74 & 37.54 & 68.19 & 68.33 & 90.41 & 90.58 & 1.45 \\ 
                                & hltbio-lg.fsrrfprf.json                                                    & 90.92 & 36.7  & 65.29 & 69.67 & 94.86 & 93.88 & 0.8  \\ \hline
\multirow{5}{*}{UAmsterdam}     & UAmsterdam\_bergen\_mistral-7b - [X].jsonl.json                     & 89.52 & 33.42 & 68.59 & 65.31 & 86.54 & 98.26 & 0.43 \\ 
                                & UAmsterdam\_bergen\_llama-8b - [X].jsonl.json                       & 87.3  & 33.17 & 76.94 & 65.04 & 74.23 & 97.95 & 0.82 \\ 
                                & UAmsterdam\_bergen\_pisco-mistral - [X].jsonl.json                  & 91.84 & 33.83 & 71.67 & 69.06 & 71.08 & 76.43 & 6.43 \\ 
                                & UAmsterdam\_bergen\_pisco-llama - [X].jsonl.json                    & 93.91 & 35.28 & 83.78 & 66.78 & 11.34 & 64.52 & 3.23 \\ 
                                & UAmsterdam\_bergen\_llama-70b - [X].jsonl.json                      & 93.12 & 37.04 & 81.31 & 67.25 & 98.84 & 98.24 & 0.7  \\ \hline
GEHC                            & task\_b\_output\_run\_gehc - [X].json.json                 & 94.55 & 37.21 & 67.29 & 63.09 & 73.59 & 94.43 & 0    \\ \hline
\multirow{8}{*}{dal}            & empd.json                                                                  & 88.01 & 37.3  & 79.75 & 68.42 & 94.58 & 83.52 & 0.5  \\ 
                                & rrf\_monot5-msmarco\_deepseek-r1.json                                      & 93.82 & 36.99 & 77.7  & 67.19 & 95.58 & 96.72 & 0.41 \\ 
                                & afmmdn.json                                                                & 94.4  & 38.08 & 81.09 & 64.96 & 88.24 & 93.49 & 0.77 \\ 
                                & rmmdn.json                                                                 & 93.23 & 36.1  & 82.33 & 63.07 & 97.73 & 96.36 & 0.4  \\ 
                                & afmmd.json                                                                 & 93.99 & 37.66 & 83.29 & 64.32 & 85.9  & 98.2  & 0.45 \\ 
                                & rmmln.json                                                                 & 93.94 & 37.68 & 88.45 & 67.26 & 99.3  & 97.75 & 1.12 \\ 
                                & rrf\_monot5-msmarco\_llama70b.json                                         & 94.72 & 37.63 & 88.7  & 66.7  & 99.36 & 97.22 & 1.11 \\ 
                                & expd.json                                                                  & 92.2  & 34.28 & 80.82 & 67.92 & 85.71 & 81.61 & 1.24 \\ \hline
\multirow{2}{*}{EvalHLTCOE}     & hltcoe-eval-svc-smoothed-sonnet - [X].jsonl.json                   & 91.79 & 38.46 & 77.37 & 68.53 & 85.26 & 87.41 & 1.36 \\ 
                                & hltcoe-eval-common-smoothed-sonnet - [X].jsonl.json                & 94.12 & 41.16 & 75.69 & 70.62 & 85.19 & 87.23 & 1.06 \\ \hline
\multirow{10}{*}{uniud}         & run7\_no\_rerank\_index-passages-dense\_Llama-3.3-70B-Instruct.json        & 79.69 & 28.95 & 43.39 & 69.17 & 68.29 & 48.68 & 9.83 \\ 
                                & run3\_no\_rerank\_index-passages-dense\_Llama-3.1-8B-Instruct.json         & 86.47 & 37.65 & 70.21 & 68.33 & 45.93 & 41.14 & 4.01 \\ 
                                & run5\_no\_rerank\_index-passages-sparse\_Llama-3.3-70B-Instruct.json       & 91.64 & 35.6  & 56.81 & 70.16 & 93.85 & 79.24 & 0.29 \\ 
                                & run4\_rerank\_index-passages-dense\_Llama-3.1-8B-Instruct.json             & 86.04 & 36.14 & 69.34 & 68.01 & 52.97 & 47.78 & 2.22 \\ 
                                & run1\_no-rerank\_index-passages-sparse\_Llama-3.1-8B-Instruct.json         & 91.31 & 35.75 & 58.04 & 66.19 & 75.63 & 77.15 & 0.75 \\ 
                                & run9\_no\_rerank\_index-passages-sparse\_gpt-4o-mini.json                  & 90.74 & 37.44 & 70.3  & 66.87 & 70.15 & 79.84 & 0.4  \\ 
                                & run10\_no\_rerank\_index-passages-dense\_gpt-4o-mini.json                  & 90.15 & 35.78 & 68.63 & 68.24 & 31.63 & 57.32 & 6.37 \\ 
                                & run8\_rerank\_index-passages-dense\_Llama-3.3-70B-Instruct.json            & 76.8  & 27.02 & 36.87 & 64.84 & 60.4  & 43.55 & 9.95 \\ 
                                & run2\_rerank\_index-passages-sparse\_Llama-3.1-8B-Instruct.json            & 86.91 & 35.39 & 55.3  & 67.62 & 80    & 77.03 & 3.04 \\ 
                                & run6\_rerank\_index-passages-sparse\_Llama-3.3-70B-Instruct.json           & 90.81 & 33.92 & 52.49 & 67.92 & 85.8  & 71.6  & 0.6  \\ \hline
\multirow{3}{*}{hltcoe-multiagt} &
  hltcoe-multiagt.llama70B.lg-w-ret\_p-nt-s\_cite - [X].jsonl.json &
  95.48 &
  39.55 &
  72.63 &
  68.18 &
  99.14 &
  97.8 &
  0.24 \\ 
                                & hltcoe-multiagt.llama70B.lg-w-ret\_lpq-nt-s\_cite - [X].jsonl.json & 93.5  & 39.45 & 70.31 & 67.18 & 96.12 & 96.31 & 0.74 \\ 
                                & hltcoe-multiagt.llama70B.ag\_sw-ret\_plq-6-wbg-limit .jsonl.json           & 96.41 & 37.09 & 71.98 & 69.16 & 96.21 & 96.23 & 0.63 \\ \hline
UDInfo                          & task\_b\_output\_reranker\_sum - [X].json.json                & 90.63 & 36    & 59.87 & 66.25 & 66.32 & 92.43 & 0.99 \\ \hline \hline
\end{tabular}%
}
\caption{\textbf{(Task B)} Evaluation of runs in \textit{early submissions} for answer and citation evaluation using the automated evaluation scheme}
\label{tab:task-b-automated-answer-eval-final}
\end{table}

\section{Results and Discussion} 

\subsection{Task A (Grounding Answer):}
The detailed results of participants' top-priority runs along with the baseline run under expert evaluation using the metrics discussed in Section \ref{sec:task-a-eval} are provided in Fig.~\ref{fig:task-a-supports-precision}, \ref{fig:task-a-supports-recall}, and \ref{fig:task-a-contadicts-precision-recall}. For the \textit{Supports} class, two runs \texttt{gehc\_htic\_task\_a} and \texttt{LLM\_BM25} achieved the highest relaxed precision score of $68.33$ while the baseline run obtained a score of $55$.
Most runs did not yield satisfactory results in identifying contradictory citations; the baseline system achieved the highest scores, $8.33$ and $10$ for precision and recall, respectively. The performance of the submitted runs, as evaluated by the automated evaluation, is shown in Table \ref{tab:task-a-automatic-eval}.  The run \texttt{LLM\_NLI\_BM25} achieved an F1-score of $67.74$ on the \textit{Supports} class, while the run \texttt{task\_a\_run2} achieved an F1-score of $15.67$ on the \textit{Contradicts} class. Both expert and automated evaluations demonstrate consistent trends indicating that systems struggle to identify contradictory citations. It also shows that our baseline system outperformed many submitted runs in identifying supporting contract terms. It indicates substantial scope for further developing methods to accurately identify citations that contradict the generated answers.

\subsection{Task B (Reference Attribution):}
We perform detailed expert evaluations for \textit{early submissions} on multiple levels for answer quality (Tables \ref{tab:res-acc}, \ref{tab:res-quality}, \ref{tab:res-recall}), citation quality (Table \ref{tab:res-citation-quality}), and document relevance (Table \ref{tab:res-doc-relevance}). These results are reported on the completed runs (\textit{cf} Table \ref{tab:expert-evaluations}). For answer accuracy (deemed acceptable for a given question), most runs achieved over 90\% accuracy except the baseline run. For precision of the answer quality, the run \texttt{UAmsterdam\_bergen\_llama-70b - [X].jsonl} achieved a maximum score of $71.55\%$, while the run \texttt{task\_b\_output\_run\_gehc - [X].json}  recorded the lowest precision of the $44.93$. For redundancy, the run \texttt{rrf\_monot5-msmarco\_llama70b} recorded the lowest redundancy score of $12.2$. Some runs produced answers with a harmful score of zero or near zero. The run \texttt{hltbio-gpt5.searcher} obtained consistently high recall scores on multiple settings compared to the other submitted runs. The results of automated metrics on \textit{early submissions} and \textit{final submissions} are shown in Table \ref{tab:task-b-automated-answer-eval-early}, and \ref{tab:task-b-automated-answer-eval-final}, respectively.
It can be observed that our automated completeness (answer evaluation) scores aligned with the expert evaluation of the recall metric. Similar observations are made for the citation evaluation as well. We will conduct a detailed analysis of the correlation between expert and automated evaluation in future work.

\section{Conclusion}
This overview of the TREC 2025 BioGen track discussed the tasks, datasets, evaluation metrics, participating systems, and their performance. We evaluated the performance of the submitted runs across multiple levels (answers, citations, and documents) using expert and automated evaluation. For Task A, most teams followed the NLI base approach. For Task B, most teams used a two-step RAG approach: first, they retrieved documents via lexical search (BM25), then re-ranked them to obtain the top-\textit{k} relevant documents/snippets. In the second stage, LLMs were used to generate the answer, citing appropriate documents. We hope that introducing the task has created ground-truth datasets to foster research on designing systems that generate answers to health-related questions with appropriate citations, thereby providing a trusted, reliable source to support the assertions in the answers. 

\section*{Acknowledgments}
This research was supported by the Intramural Research Program of the National Institutes of Health (NIH). The contributions of the NIH authors are considered Works of the United States Government. The findings and conclusions presented in this paper are those of the authors and do not necessarily reflect the views of the NIH or the U.S. Department of Health and Human Services. This research was further supported by the U.S. National Library of Medicine (NLM) funded projects R01LM011934 (WH, SB, KR) and R01LM014508 (KR). The authors thank Srishti Kapur of Centaur AI for managing the expert evaluation process. This work utilized the computational resources of the NIH HPC Biowulf cluster (\url{https://hpc.nih.gov}). 

\bibliographystyle{unsrt}
\bibliography{sample}

\begin{thebibliography}{10}

\bibitem{zhao2024heterogeneous}
Wenting Zhao, Zhongfen Deng, Shweta Yadav, and Philip~S Yu.
\newblock Heterogeneous knowledge grounding for medical question answering with retrieval augmented large language model.
\newblock In {\em Companion Proceedings of the ACM on Web Conference 2024}, pages 1590--1594, 2024.

\bibitem{wu2024well}
Kevin Wu, Eric Wu, Ally Cassasola, Angela Zhang, Kevin Wei, Teresa Nguyen, Sith Riantawan, Patricia Shi~Riantawan, Daniel~E Ho, and James Zou.
\newblock How well do llms cite relevant medical references? an evaluation framework and analyses.
\newblock {\em arXiv e-prints}, pages arXiv--2402, 2024.

\bibitem{basaragin-etal-2024-know}
Bojana Ba{\v{s}}aragin, Adela Ljaji{\'c}, Darija Medvecki, Lorenzo Cassano, Milo{\v{s}} Ko{\v{s}}prdi{\'c}, and Nikola Milo{\v{s}}evi{\'c}.
\newblock How do you know that? teaching generative language models to reference answers to biomedical questions.
\newblock In Dina Demner-Fushman, Sophia Ananiadou, Makoto Miwa, Kirk Roberts, and Junichi Tsujii, editors, {\em Proceedings of the 23rd Workshop on Biomedical Natural Language Processing}, pages 536--547, Bangkok, Thailand, August 2024. Association for Computational Linguistics.

\bibitem{jamiaocae014}
William Hersh.
\newblock {Search still matters: information retrieval in the era of generative AI}.
\newblock {\em Journal of the American Medical Informatics Association}, 31(9):2159--2161, 01 2024.

\bibitem{phan2021scifive}
Long~N. Phan, James~T. Anibal, Hieu Tran, Shaurya Chanana, Erol Bahadroglu, Alec Peltekian, and Grégoire Altan-Bonnet.
\newblock Scifive: a text-to-text transformer model for biomedical literature, 2021.

\bibitem{touvron2023llama}
Hugo Touvron, Louis Martin, Kevin Stone, Peter Albert, Amjad Almahairi, Yasmine Babaei, Nikolay Bashlykov, Soumya Batra, Prajjwal Bhargava, Shruti Bhosale, et~al.
\newblock Llama 2: Open foundation and fine-tuned chat models.
\newblock {\em arXiv preprint arXiv:2307.09288}, 2023.

\bibitem{attal2023dataset}
Kush Attal, Brian Ondov, and Dina Demner-Fushman.
\newblock A dataset for plain language adaptation of biomedical abstracts.
\newblock {\em Scientific Data}, 10(1):8, 2023.

\bibitem{yadav2022chq}
Shweta Yadav, Deepak Gupta, and Dina Demner-Fushman.
\newblock Chq-summ: A dataset for consumer healthcare question summarization.
\newblock {\em arXiv preprint arXiv:2206.06581}, 2022.

\bibitem{gupta2024empowering}
Deepak Gupta and Dina Demner-Fushman.
\newblock Empowering language model with guided knowledge fusion for biomedical document re-ranking.
\newblock In {\em International Conference on Artificial Intelligence in Medicine}, pages 251--260, 2024.

\bibitem{demner2007answering}
Dina Demner-Fushman and Jimmy Lin.
\newblock Answering clinical questions with knowledge-based and statistical techniques.
\newblock {\em Computational Linguistics}, 33(1):63--103, 2007.

\bibitem{gao2023enabling}
Tianyu Gao, Howard Yen, Jiatong Yu, and Danqi Chen.
\newblock Enabling large language models to generate text with citations.
\newblock In {\em Proceedings of the 2023 Conference on Empirical Methods in Natural Language Processing}, pages 6465--6488, 2023.

\bibitem{gupta2026bioaceautomatedframeworkbiomedical}
Deepak Gupta, Davis Bartels, and Dina Demner-Fushman.
\newblock Bioace: An automated framework for biomedical answer and citation evaluations.
\newblock 2026.

\bibitem{bartels-etal-2025-large}
Davis Bartels, Deepak Gupta, and Dina Demner-Fushman.
\newblock Can large language models accurately generate answer keys for health-related questions?
\newblock In Wanxiang Che, Joyce Nabende, Ekaterina Shutova, and Mohammad~Taher Pilehvar, editors, {\em Proceedings of the 63rd Annual Meeting of the Association for Computational Linguistics (Volume 2: Short Papers)}, pages 354--368, Vienna, Austria, July 2025. Association for Computational Linguistics.

\bibitem{biogen2024TREC}
Deepak Gupta, Dina Demner-Fushman, William Hersh, Steven Bedrick, and Kirk Roberts.
\newblock {Overview of TREC 2024 Biomedical Generative Retrieval (BioGen) Track}.
\newblock In {\em {The Thirty-Third Text REtrieval Conference Proceedings (TREC 2024)}}, {NIST Special Publication}. {National Institute of Standards and Technology (NIST)}, 2024.

\end{thebibliography}

\end{document}